\RequirePackage{fix-cm}
\documentclass[smallcondensed,final,a4paper]{svjour3}       
\smartqed  
\usepackage{graphicx}

\usepackage{cite}
\usepackage{amsmath,amssymb,amsfonts}
\usepackage{textcomp}
\usepackage{algorithm}
\usepackage{algpseudocode}
\usepackage{multirow}
\usepackage{epstopdf}
\usepackage{makecell}
\usepackage{rotating}

\usepackage[font=small,labelfont=bf]{caption}

\title{\LARGE \bf
Network Coding-based Routing and Spectrum Allocation in Elastic Optical Networks for Enhanced Physical Layer Security
}
\author{*Giannis Savva, Konstantinos Manousakis, Georgios Ellinas
\thanks{This work has been partially supported by the European Union's Horizon 2020 research and innovation programme under grant agreement No 739551 (KIOS CoE) and from the Government of the Republic of Cyprus through the Directorate General for European Programmes, Coordination and Development. It was also partially supported by the Cyprus Research and Innovation Foundation under project CULTURE/AWARD-YR/0418/0014 (REALFON).}
\thanks{Giannis Savva, Konstantinos Manousakis and Georgios Ellinas are with the Department of Electrical and Computer Engineering and the KIOS Research and Innovation Center of Excellence, University of Cyprus, Nicosia 1678, Cyprus}%
\thanks{*savva.giannis@ucy.ac.cy, manousakis.konstantinos@ucy.ac.cy, gellinas@ucy.ac.cy}%
}

\begin{document}

\maketitle
\thispagestyle{empty}
\pagestyle{empty}

\begin{abstract}

In this work, an eavesdropping-aware routing and spectrum allocation approach is proposed utilizing network coding (NC) in elastic optical networks (EONs). To provide physical layer security in EONs and secure the confidential connections against eavesdropping attacks using NC, the signals of the confidential connections are combined (XOR-ed) with other signals at different nodes in their path, while transmitted through the network. The combination of signals through NC significantly increases the security of confidential connections, since an eavesdropper must access all individual signals, traversing different links, in order to decrypt the combined signal. A novel heuristic approach is proposed,  that solves the combined network coding and routing and spectrum allocation (NC-RSA) problem, that also takes into account additional NC constraints that are required in order to consider a confidential connection as secure. Different routing and spectrum allocation strategies are proposed, aiming to maximize the level of security provided for the confidential demands, followed by an extensive performance evaluation of each approach in terms of the level of security provided, as well as the spectrum utilization and blocking probability, under different network parameters. Performance results demonstrate that the proposed approaches can provide efficient solutions in terms of network performance, while providing the level of security required for each demand. 

\end{abstract}


\section{INTRODUCTION}

Elastic optical networks (EONs) provide improved transmission performance and more flexible spectrum allocation compared to conventional WDM optical networks. In EONs, the C-band can be separated in slices (frequency slots) of $25$, $12.5$, and $6.25$ GHz. Subsequently, a connection can be allocated to a number of frequency slots to satisfy its required bit rate. To establish a connection in EONs, a viable solution to the RSA problem must be found, which consists of finding a path (routing - R) from the source to the destination, along with the allocation of the required spectrum slots (spectrum allocation - SA). To find a feasible RSA solution, three basic constraints must be satisfied: (i) the \textit{spectrum continuity} constraint, where each demand must be allocated the same frequency slots on each link of the selected path, (ii) the \textit{non-overlapping} constraint, where a frequency slot can only be allocated to one demand at a time, and (iii) the \textit{spectrum contiguity} constraint, where the slots serving each demand must be contiguous~\cite{EONs1,RSA}. 

Optical networks are considered one of the most important critical infrastructures, as many services (e.g., cloud computing) and critical applications (e.g., national security, public health) rely on these networks for data exchange.   Due to the type and the amount of data transmitted, the security of optical networks is of paramount importance, since even attacks over a short time period can still compromise large amounts of critical data.

There are numerous real-world scenarios where optical networks were breached (e.g., utilizing tapping devices to access the signal transmitted via optical fibers). Hence, optical layer security has received considerable attention in recent years. Optical layer security can be divided into different categories based on the type and the purpose of the threat. Such security threats include the observation of the existence of communication (privacy), the unauthorized use of spectrum (authentication), the manipulation or destruction of data (integrity), the denial of service (availability), and the unauthorized access to information (confidentiality)~\cite{Sec1,Sec3,Furdek16,Sec2,Savva182,JAM1,SavvaJocn,Bei18,Singh16}. The focus of this work is on confidentiality, where an adversary tries to and can have undetected access to confidential data from an optical communication channel (i.e., eavesdropping) for a prolonged period of time.

Network coding (NC) is a technique where data of different connections are combined and transmitted through the network~\cite{Manley10}. In this work, NC is used to provide security in EONs against eavesdropping attacks by encrypting the confidential data. Thus, with the use of NC, it is extremely difficult for an adversary to compromise any confidential connection, since a number of different connections that traverse different routes will have to be compromised, in order for the attacker to make sense of the accessed confidential data. This work significantly extends the concept presented in~\cite{Savva19} by providing an extended analysis of NC within optical networks, followed by novel policies for allocating a path and spectrum slots for the confidential demands. Further, additional properties are considered for the paths used in the XOR operations, which are subsequently incorporated within the heuristic algorithms. Finally, extensive simulations are performed to compare all policies in terms of spectrum utilization, blocking probability, and the level of security provided for the confidential demands.

The rest of the paper is organized as follows. The general concept of network coding and its implementations in optical networks are discussed in Section~\ref{coding}. The combined NC-RSA problem is described in Section~\ref{problem}, followed by the proposed heuristic algorithm in Section~\ref{heuristic}. Next, the performance results are presented in Section~\ref{results}, while Section~\ref{conclusion} offers some concluding remarks.

\section{Network Coding Implementations in Optical Networks}
\label{coding}

\begin{figure}[htbp]
\centerline{\includegraphics[scale = 0.25]{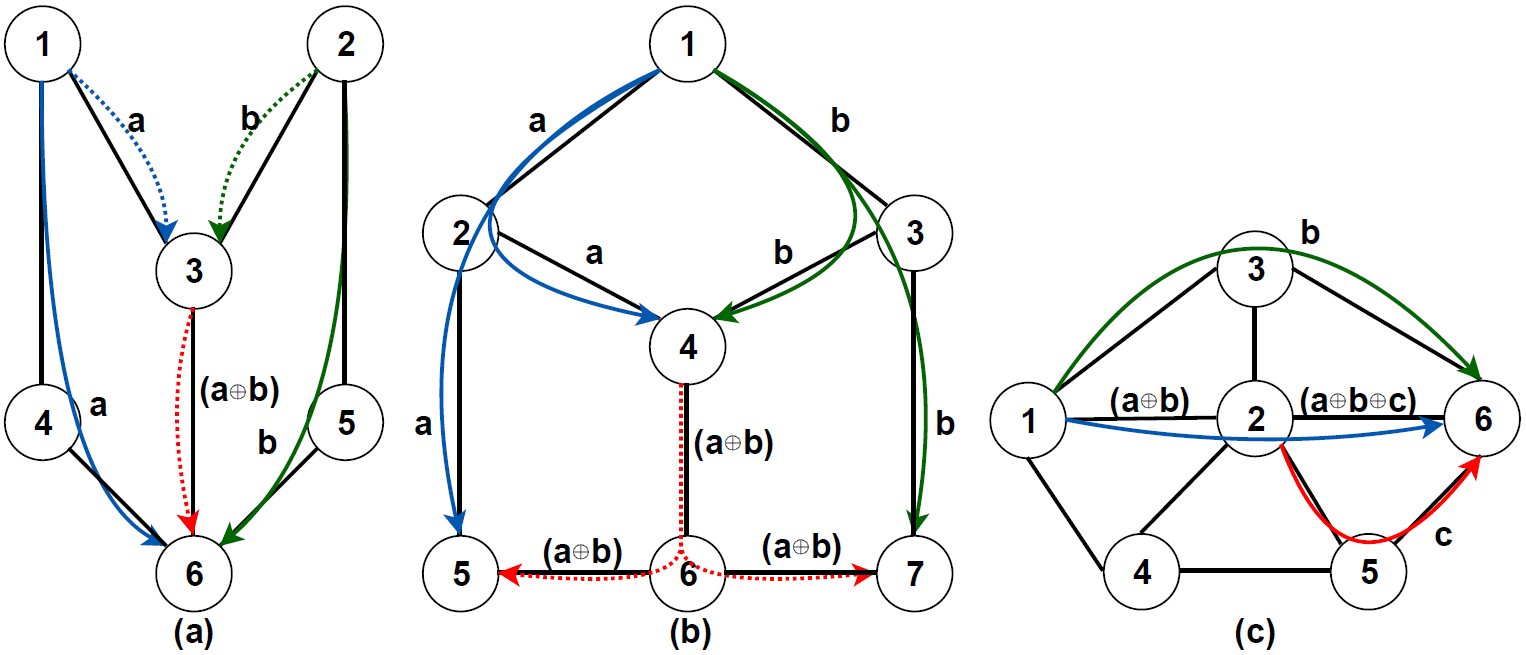}}
\caption{Utilizing NC for (a) protection, (b) multicasting, and (c) security.}
\label{NC_example}
\end{figure}

The application of NC to optical networks from the algorithmic and infrastructural perspectives has been investigated by the authors in~\cite{Manley10}. Furthermore, using NC in conjunction with the RSA process, several connections can combine their transmitted data, enabling protection, multicasting, and security functionalities within the optical network. In the following, a short description of how these functionalities can be achieved, along with a discussion on the physical implementation of NC in optical networks are presented.\\
\\
\noindent {\bf (a) NC-based Protection in Optical Networks}\\
NC-based protection in optical networks has been widely studied~\cite{Manley10,Ramirez14,Kamal14,Hai17}. For such an approach, intermediate network nodes process a set of signals such that a destination node will receive a linear combination of signals over several disjoint paths. Then, in the case of a failure in the primary path, the destination node will still be able to recover the failed signal from the rest of the received combined signals arriving at the destination from paths that are not affected by the failure. For instance, consider the network shown in Fig.~\ref{NC_example}(a). In this example, connections $1-4-6$ and $2-5-6$ transmit datastreams $a$ and $b$, respectively. To provide protection for the primary paths, using network coding, source nodes $1$ and $2$ also transmit their datastreams to node $3$, which performs an XOR operation ($a \oplus b$). Subsequently, the resulting signal ($a \oplus b$) is transmitted to the destination (i.e., node $6$). In the case that connection $1-4-6$ fails, datastream $a$ can still be recovered by the destination node as $a = (a \oplus b) \oplus b$. Similarly, in the case that path $2-5-6$ fails, datastream $b$ can still be obtained as $b = (a \oplus b) \oplus a$.

\noindent {\bf (b) NC-based Multicasting in Optical Networks}\\
NC can also be used for multicasting in optical networks. Authors in~\cite{Manley10,Agarwal04,Yang16} demonstrate the utilization of network coding for multicasting connections to improve network throughput. The concept of NC in multicasting is to divide the original traffic into different lower-rate sub-streams and design trees in such a way that the destination nodes will be able to decode the incoming signals based on specific XOR operations. In Fig.~\ref{NC_example}(b), assume that node $1$ is the source node and needs to transmit  datastreams $a$ and $b$ to both destination nodes $5$ and $7$. One possible solution using network coding is to transmit datastream $a$ to node $5$ and datastream $b$ to node $7$, perform the XOR operation $(a \oplus b)$ at node $4$, and transmit the resulting datastream $(a \oplus b)$ to both destination nodes through paths $4-6-7$ and $4-6-5$. Hence, each node can obtain both $a$ and $b$ by using the received signals. For example, node $5$ receives datastream $a$ and $(a \oplus b)$. Then, $b$ can be obtained through the process $b = a \oplus (a \oplus b)$. A similar procedure can also be performed at node $7$ to obtain datastreams $a$ and $b$~\cite{Kim09,Yang16}.\\
\\
\noindent {\bf (c) NC-based Security in Optical Networks}\\
Authors in~\cite{Engel16} studied the effectiveness of linear network coding (LNC) in optical networks to protect connections from security threats such as eavesdropping and jamming attacks. To enhance the network with security capabilities, a confidential connection can transmit a coded version of its data. Hence, in the case of an eavesdropping attack, the attacker must have knowledge of both the coded version of the signal and all connections used to perform the XOR operations in order to access the confidential data. Recently, work in~\cite{Savva19} investigated the problem of network security in elastic optical networks using NC through a heuristic approach. This is also the main topic of this work, in which a significant extension of the approach presented in~\cite{Savva19} is provided.

Fig.~\ref{NC_example}(c) presents a simple example where network coding can be used to provide security for a confidential connection in an optical network. As shown in this figure, the confidential datastream $a$ is transmitted using path $1-2-6$. Also, two non-confidential connections are established in the network using paths $1-3-6$ and $2-5-6$, and transmit datastreams $b$ and $c$, respectively. To provide security for the confidential connection using path $1-2-6$, source node $1$ transmits a coded version of datastream $a$, namely $a \oplus b$, over link $1-2$, and additional security is provided at link $2-6$ by also using connection $2-5-6$ and datastream $c$. Thus, ($a\oplus b \oplus c$) is transmitted at the second link of the confidential path. To acquire the confidential datastream $a$ at node $6$, the receiver can perform the operation $a=b \oplus c \oplus (a \oplus b \oplus c)$. Hence, an eavesdropper must also gain access to both datastreams $b$ and $c$ in the network in order to be able to make sense of the confidential datastream $a$. In other words, the eavesdropper must simultaneously tap $3$ different links at different network locations to compromise datastream $a$.\\
\\
\noindent {\bf (d) Physical Layer Implementation of NC in Optical Networks}\\
In opaque networks, NC can be easily implemented, since signals can be received, combined, and transmitted by the intermediate nodes, providing an encrypted version of the signal at each link. However, for such a technique to work in transparent optical networks, the nodes must be equipped with additional hardware in order to perform the XOR operation at the physical (optical) layer. Furthermore, all involved signals must use the same spectrum resources in order for network coding to be enabled. It was shown that NC in all-optical networks can be realized by all-optical XOR logic gates. Specifically, a review of an optical implementation of all-optical XOR gates was performed in~\cite{Zhang05}. All-optical XOR gates are typically based on semiconductor optical amplifiers (SOAs) that offer low-power consumption, easy deployment, and short-latency. Experimental works have also demonstrated NC using XOR operations that can be performed at line speed for transmission above $10$ Gbps and up to $100$ Gbps with different modulation formats~\cite{Kong13,Chen15,Lu16}. In addition, authors in~\cite{Li14} proposed and experimentally demonstrated a coherent optical layer network coding scheme on polarization-multiplexed differential quaternary phase-shift keying (PM-DQPSK) signals, in which the two PM-DQPSK signal components in the network-coded signal occupy the same channel and cannot be separated by conventional demultiplexing means. Finally, authors in~\cite{Liu13} demonstrated, using OFDM signals, that the NC technique does not require symbol-level synchronization. Thus, the feasibility of optical NC was clearly demonstrated and is a technology that could be implemented in EONs to provide security at the physical layer against eavesdropping attacks.

\section{Problem Description} \label{problem}

To provide security for confidential connections in the network against eavesdropping attacks, the combined NC-RSA problem must now be solved. More specifically, in order to find the path and the spectrum required for a confidential connection, the path is chosen so as to take advantage of the existing connections established in the network, that are utilized to perform a number of XOR operations in the nodes that traverse the confidential path. Further, it should be noted that the spectrum slots selected for the confidential connection can significantly affect the choice of existing connections that can be involved in the XOR process. To consider a confidential connection as secure when using network coding, the following constraints must be satisfied.

\begin{itemize}
    \item \textbf{Encrypted Transmission (ET)}: All links of the path allocated to a confidential connection must transmit an encrypted version of their data with at least one XOR operation with other currently established connections in the network.
    \item \textbf{Frequency Slot Matching (FSM)}: At least a subset of the frequency slots utilized by the confidential connection must have the same id with the slots of the rest of the already established connections used in the XOR operations.
\end{itemize}

To satisfy the \textbf{ET} constraint, an existing connection must have at least two common nodes with the confidential connection (the first node will be used to encrypt the confidential connection and the second node will be used to decrypt it). Thus, an existing connection with at least two common nodes with the confidential connection can either be used to provide security for the entire path of the confidential demand (i.e., when the source and destination nodes are common nodes), or it can be used to provide security for a segment of the confidential connection (i.e., source/intermediate to intermediate/destination node). Also, in this work, the order in which the common nodes are traversed by the connections involved in the XOR process must be the same, which will also ensure that the direction of the connections traversing the nodes are the same. Finally, the \textbf{ET} constraint ensures that for a secure confidential connection, the selected existing connections must collectively secure all links of the path allocated to that connection.

To satisfy the \textbf{FSM} constraint, at least a subset of the frequency slots utilized by the confidential connection must have the same id with the frequency slots of the rest of the existing connections used in the XOR operation. This is the case, since it is assumed that no frequency conversion is performed at intermediate nodes, and therefore, the signals used for the XOR operation must be on the same frequency. However, the confidential connection is considered as secure even when only part of the signal (a subset of the frequency slots) is XOR-ed, since the eavesdropper would still have to access all connections used in the encryption process in order to decrypt and make sense of the accessed confidential data.

Finally, it is important to note that to satisfy both constraints, all connections involved in the encryption process (i.e., both the confidential and the currently established connections used in the XOR operation) must be link-disjoint. This is the case, since the connections used in the XOR operation must transmit signals that are at least partially utilized on the same frequency. Thus, considering also the \textit{non-overlapping} constraint of the RSA problem, all paths involved in this process must be link-disjoint.

\begin{figure}[htbp]
\centerline{\includegraphics[scale = 0.15]{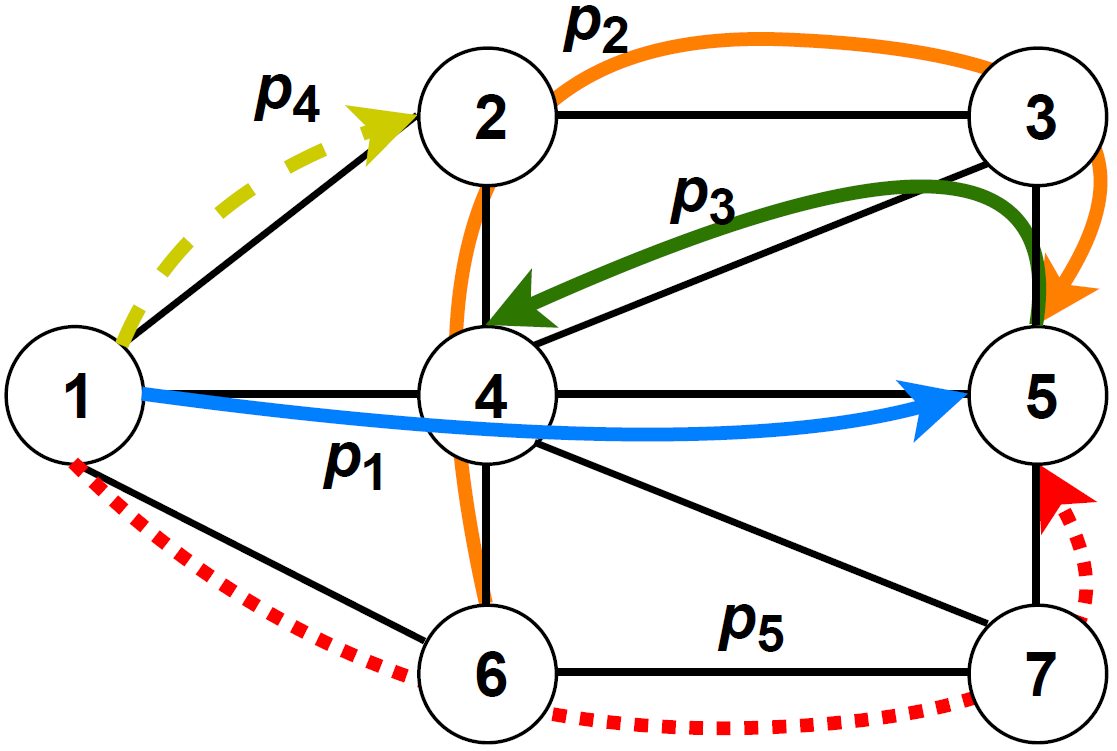}}
\caption{Example of network coding for the establishment of a secure confidential connection.}
\label{network_coding}
\end{figure}

The following example (Fig.~\ref{network_coding}) provides an illustration of the aforementioned constraints in combination with the establishment of a secure confidential connection. In this example, $4$ connections ($p_2,p_3,p_4$, $p_5$) are currently established in the network, provisioned as shown in Table~\ref{ex1_t1}, where each connection, $p_x$, transmits datastream $c_x$. Further, a confidential connection $p_1$ must be established in the network, requiring $2$ spectrum slots and using path $1-4-5$.

As also discussed previously, only the existing connections with two common nodes with the confidential connection can be used to provide security for $p_1$ using NC. Thus, $p_4$ and $p_3$ cannot be used to provide security for $p_1$, as $p_4$ does not have two common nodes with $p_1$ and $p_3$ traverses its common nodes with $p_1$ in the opposite order compared to $p_1$. Further, the group of spectrum slots that will be assigned for connection $p_1$ is very important, since at least a subset of the frequency slots utilized by the confidential connection as well as the already existing connections that take part in the XOR process must be the same (\textbf{FSM} constraint). Thus, if spectrum slots $4-5$ were used for $p_1$, then only $p_5$ can be used to provide security for the connection, while if spectrum slots $2-3$ are used, then both $p_2$ and $p_5$ can be used to provide (a higher level of) security for $p_1$. Table~\ref{ex1_t2} presents the transmitted datastreams for each link of the confidential connection, if allocated using spectrum slots $2-3$. As shown in Table~\ref{ex1_t2}, an encrypted version of the confidential signal is transmitted at all links of the path, satisfying also the \textbf{ET} constraint. In this example, the destination node can successfully decrypt the data, whereas the eavesdropper must gain access to multiple additional connections that are allocated at link-disjoint paths (i.e., $p_2$ and $p_5$) apart from the confidential one, in order to make sense of the accessed confidential datastream.

\begin{table}[htbp]
\parbox{.60\linewidth}{
\caption{Paths and spectrum slots allocated to existing connections in the network.}
\centering
\begin{tabular}{|c|c|c|}
\hline
\textbf{\begin{tabular}[c]{@{}c@{}}Connection\\ Id\end{tabular}} & \textbf{\begin{tabular}[c]{@{}c@{}}Path\\ Selected\end{tabular}} & \textbf{\begin{tabular}[c]{@{}c@{}}Spectrum\\ Slots Allocated\end{tabular}} \\ \hline
$p_2$ & $6 - 4 - 2 - 3 - 5$ & $1 - 3$ \\ \hline
$p_3$ & $5 - 3 - 4$ & $1 - 4$ \\ \hline
$p_4$ & $1 - 2$ & $1 - 2$ \\ \hline
$p_5$ & $1 - 6 - 7 - 5$ & $2-5$ \\ \hline
\end{tabular}
\label{ex1_t1}
}
\hfill
\parbox{.30\linewidth}{
\caption{Datastream transmitted at each link of the confidential connection $p_1$.}
\centering
\begin{tabular}{|c|c|}
\hline
\textbf{Link} & \textbf{\begin{tabular}[c]{@{}c@{}}Transmitted\\ datastream\end{tabular}} \\ \hline
$1 - 4$ & $c_1 \oplus c_5$ \\ \hline
$4 - 5$ & $c_1 \oplus c_2 \oplus c_5$ \\ \hline
\end{tabular}
\label{ex1_t2}
}
\end{table}


\section{Proposed Heuristic Algorithm}
\label{heuristic}
To solve the combined NC-RSA problem, the proposed heuristic algorithm is divided into the routing (R) and spectrum allocation (SA) sub-problems. A network planning scenario is considered, where all demands are known a priori and each demand is described by a $4$-tuple (\textit{s,d,B,c}), denoting the source, destination, bit-rate, and confidentiality, respectively. Confidentiality in this case is defined as a binary variable which describes the demand as confidential ($1$) or non-confidential ($0$). It is noted that different classes of confidential connections could be utilized based on the minimum threshold of XOR operations (designated as $T$ in this work) required per link. However, in this work, only the case of confidential ($T=1$) and non-confidential demands will be considered.

\subsection{Pre-processing Phase}
Prior to the utilization of the heuristic approach, a pre-processing phase is required that pre-calculates the candidate paths for each $s$-$d$ pair. Also, a node-matching procedure takes place, to ascertain whether two paths have at least two common nodes traversed in the same order, so as to satisfy the \textbf{ET} and \textbf{FSM} constraints, that are required for a feasible NC-RSA solution for the confidential demands. 

First, a set of $k$ candidate paths is pre-calculated for each $s$-$d$ pair. Subsequently, to implement NC, the security level that each candidate path can provide to the rest of the candidate paths must be calculated. According to the \textbf{ET} constraint, all links traversed by a secure confidential connection must transmit an encrypted version of the data. This means that at least one XOR operation with the established connections must be performed for all links in the path used by the confidential connection. Let $P$ be the set of all candidate paths, $p$ the path under consideration ($p \in P$), and $P'$ the set of the rest of the candidate paths that may be used by path $p$ to provide encryption over specific links of path $p$ ($P = P' \cup p$). Also, let $\delta_{p'l}^{p}$ be a coefficient which is equal to $1$ if paths $p\in P$ and $p'\in P'$ have two common nodes and the $l^{th}$ link of path $p$ can be ``covered'' (i.e., encryption is provided) by path $p'$, and $0$ otherwise. 

Figure~\ref{net_cod_delta} provides an example of the $\delta$ coefficients for $3$ paths $\{3-1-2-4, 2-5-6-4, 2-3-5\}$ utilized by $3$ connections in the network. Hence, the links that are covered for each pair of connections must be calculated ($\delta$ coefficient), with regards to the common nodes of each pair of paths.

\vspace{0.1in}

\begin{minipage}{\textwidth}
  \begin{minipage}[b]{0.49\textwidth}
    \centerline{\includegraphics[scale = 0.35]{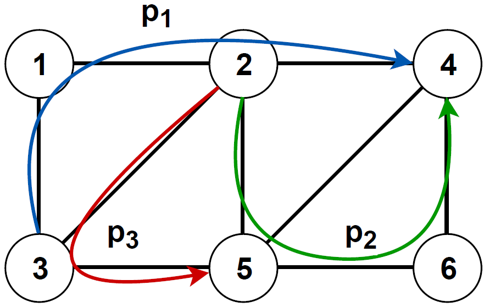}}
    \captionof{figure}{Example of $\delta$ coefficient calculation \newline for $3$ connections established in the network.}
    \label{net_cod_delta}
  \end{minipage}
\hspace{0.25in}  
  \begin{minipage}[b]{0.30\textwidth}
    \captionof{table}{$\delta$ coefficients for $3$ paths established in the network.}
    \begin{tabular}{|c|c|c|c|}
\hline
$\delta_{p'l}^{p}$ & $\delta_{p'1}^{p}$  & $\delta_{p'2}^{p}$  & $\delta_{p'3}^{p}$  \\ \hline
$\delta_{2l}^{1}$ & 0 & 0 & 1 \\ \hline
$\delta_{3l}^{1}$ & 0 & 0 & 0 \\ \hline
$\delta_{3l}^{2}$ & 1 & 0 & 0 \\ \hline
$\delta_{1l}^{2}$ & 1 & 1 & 1 \\ \hline
$\delta_{1l}^{3}$ & 0 & 0 & - \\ \hline
$\delta_{2l}^{3}$ & 1 & 1 & - \\ \hline
\end{tabular}
\label{ex_2tb3}
          \end{minipage}
  \end{minipage}

\vspace{0.2in}



Assume that in this case initially the path under consideration is path $p_2$. To check whether the rest of the connections can provide security for the connection using path $p_2$,the following procedure is followed: path $p_2$ has two common nodes with path $p_1$ (nodes $2$ and $4$) and since these nodes are the source and destination nodes of $p_2$, then connection $p_1$ could be used to provide security for all links that belong to $p_2$. Hence, the values of $\delta_{11}^{2}$, $\delta_{12}^{2}$, and $\delta_{13}^{2}$ are equal to $1$ for the first, second, and third link of path $p_2$, respectively. It is important to note that on the other hand, connection $p_2$ can only provide security for the third link (link $2-4$) of connection $p_1$, resulting in $\delta_{23}^{1}$ = $1$. Thus, clearly, $\delta_{1l}^{2} \neq \delta_{2l}^{1}$. Similarly, the first link of $p_2$ can be secured by $p_3$, since nodes $2$ and $5$ are common between these paths and therefore $\delta_{21}^{3}$ = $1$. Also, in this case, $p_2$ can be used to provide security for all links in $p_3$, since $2$ and $5$ are the source and destination nodes of $p_3$, respectively. Therefore, $\delta_{31}^{2}$ = $\delta_{32}^{2} =1$. Table \ref{ex_2tb3} presents the $\delta$ coefficients for all path combinations in this example. 

It is noted that the order of the nodes in the path and the direction of the path itself are taken into consideration in the determination of the $\delta$ coefficient. For example, $p_3$ has two common nodes with $p_1$ (nodes $2$ and $3$), but it cannot provide any security for that connection, since in path $p_3$, node $2$ is traversed before node $3$, whereas the opposite is the case with path $p_1$. Algorithm~\ref{alg_delta} below is used to calculate the $\delta$ coefficients. Specifically, to calculate the $\delta$ coefficients, the algorithm takes as input the set of all the pre-calculated candidate paths $P$. Then, for each pair of candidate paths $p$ and $p'$, their common nodes are found and they are stored in set $CN$. Next, if the size of set $CN$ is greater than $1$ (i.e., there are at least two common nodes between paths $p$ and $p'$), then starting from the source node, for each node $i$ in path $p$, if that node is included in $CN$ (i.e., $i$ is a common node for both $p$ and $p'$), the order of the node $p(i)$ in $p'$ is stored in $S_{node}$. Subsequently, starting from the destination node of $p$, for each node $j$ that is traversed after node $i$ in $p$, if the node is included in $CN$, the order of the node $p(j)$ in $p'$ is found and is stored in $E_{node}$. If $E_{node}$ is greater than $S_{node}$ (i.e., node $p(j)$ is traversed after node $p(i)$ in both $p$ and $p'$), then all the links between nodes $p(i)$ and $p(j)$ in $p$ can be secured by path $p'$. Thus, the $\delta_{p'}^{p}$ coefficient is equal to $1$ for all these links. 

\begin{algorithm}[htbp]
  \caption{Calculation of $\delta$ coefficients} 
  \hspace*{\algorithmicindent} \textbf{Input:} \textit{Set of candidate paths P}\\
  \hspace*{\algorithmicindent} \textbf{Output:} $\delta$ coefficients
  \label{alg_delta}
  \begin{algorithmic}[1]
  	\For{each candidate path $p \in P$}
  	    \State $Z$: number of nodes $\in p$
  	    \For{each candidate path $p' \in P'$}
  	    \State Initialize empty set $CN$, 
  	    \State Find common nodes between paths $p$ and $p'$ and store them in $CN$
  	        \If {$|CN|>1$}
  	            \For{i = 1: $Z$}
  	                \If{$p(i) \in CN$}
  	                    \State $S_{node}$ = order of node $p(i)$ in $p'$
  	                    \For{j = $Z$: i+1}
  	                        \If{$p(j) \in CN$}
  	                            \State $E_{node}$ = order of node $p(j)$ in $p'$
  	                            \If{$E_{node} > S_{node}$}
  	                                \State $L'$: set of links between nodes $p(i)$ and $p(j)$ in path $p$
  	                                \State $\delta_{p'l}^{p} = 1$ $\forall l \in L'$
  	                                \State Go to next candidate path $p'$ (\textbf{line 3})
  	                            \EndIf
                            \EndIf
                        \EndFor
                    \EndIf
                \EndFor
            \EndIf
        \EndFor
    \EndFor
\end{algorithmic}
\end{algorithm}

It is important to note that, using this process, it is ensured that node $p(i)$ is traversed prior to node $p(j)$, which is also the case for path $p'$, in order for path $p'$ to provide security for path $p$. Further, if more than $2$ common nodes are found, the links secured may vary depending on which two common nodes are selected. However, following this process, the set of common nodes that result in $p'$ securing the largest number of links of path $p$ is selected. This is a direct consequence of the way the common nodes are selected (i.e., the first common node is found while searching from the source node to the destination and the second common node is found by starting from the destination node). 

\subsection{Routing}
For the routing sub-problem, the pre-calculated $k$-shortest candidate paths that are able to satisfy a requested connection are considered. These $k$-shortest paths can be sorted based on different metrics (e.g., number of hops, modulation format, most/least used nodes/links, etc.). In this work, each node ($n$) and link ($l$) are characterized by the number of established connections that use them (denoted as $U_{n}$ and $U_{l}$, respectively). Further, each path takes the value of the node/link with the highest value of $U_{n}$/$U_{l}$ among all the nodes/links it traverses ($U_{p,n}, U_{p,l}$). The paths of the {\bf non-confidential connections} can then be sorted based on the following  criteria, which provide the best results in terms of spectrum utilization and level of security, as presented in~\cite{Savva19}:

\begin{itemize}
    \item \textbf{Most used nodes/links (MUN/MUL)}: The candidate paths for each $s$-$d$ pair are sorted in descending order based on the value $U_{p,n}$/$U_{p,l}$ of each path. By selecting the path which comprises of the most used nodes/links, the number of XOR operations for a confidential connection will potentially increase.
    \item \textbf{Maximum spectrum efficiency (MSE)}: The candidate paths are sorted in descending order based on the number of hops and the modulation format used (hybrid metric defined in~\cite{ICC_Giannis}). Thus, connections are established having as an aim to maximize the spectrum efficiency of the network, rather than maximizing the number of XOR operations. 
\end{itemize}

After sorting the paths, the non-confidential connections are established using the first path that has available spectrum resources, based on one of the aforementioned sorting strategies. 

For the {\bf confidential connections}, the candidate path that produces the most XOR operations is used, which also depends on the spectrum slots selected. To achieve this, an XOR spectrum slot metric (XOR-SSM$_p$) is used, that counts the number of XOR operations performed for a specific path $p$ and group of frequency slots. In addition, the XOR-SSM$_p$ metric can be further differentiated as follows:
\begin{itemize}
\item \textbf{Minimum XOR spectrum slot metric (MXOR-SSM$_p$)}: The MXOR-SSM$_p$ metric counts the minimum number of XOR operations over all links of candidate path $p$. This is the metric that was first introduced and used in~\cite{Savva19}. By using this metric the aim is to maximize the level of security provided for the ``weakest link'' of the candidate path (the link with the least number of XOR operations performed).
\item \textbf{Average XOR spectrum slot metric (AXOR-SSM$_p$)}: The AXOR-SSM$_p$ metric calculates the average number of XOR operations per link that are provided for candidate path $p$. By using this metric the aim is to maximize the average number of XOR operations performed on each link of the candidate path (provided that the minimum threshold is met, as explained below).
\end{itemize}

\subsection{Spectrum Allocation}
\label{sal}
For the spectrum allocation sub-problem, available spectrum resources must be allocated for a requested connection satisfying the slot \textit{continuity}, \textit{contiguity}, and \textit{non-overlapping} constraints~\cite{RSA}. To calculate the number of frequency slots $f_{dp}$ required for a given connection $d$ if candidate path $p$ is selected, we use $f_{dp} = \frac{B_d}{B_{rate} \cdot MF_{p}}$, where $B_d$ is the bit-rate requested by connection $d$, $B_{rate}$ is the baud rate of each spectrum slot in the network, and $MF_{p}$ is the modulation format (expressed in bits/symbol) that can be used for candidate path $p$, which is based on the overall distance of path $p$.

\begin{figure}[htbp]
\centerline{\includegraphics[scale = 0.15]{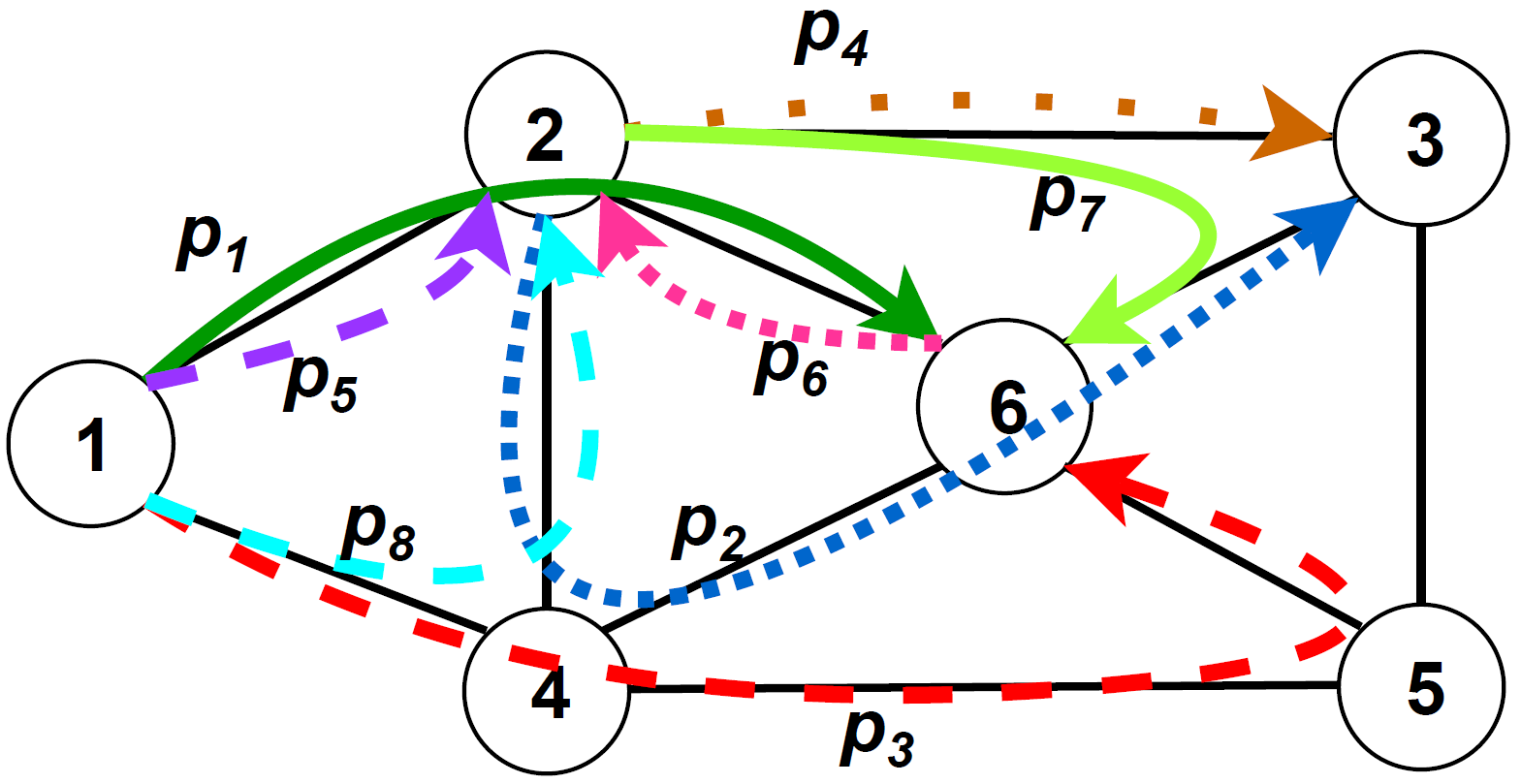}}
\caption{Example of NC-RSA algorithm for a given confidential connection.}
\label{temp_ex}
\end{figure}

For the non-confidential connections, the spectrum allocation is performed in a first-fit manner, where the first group of frequency slots from the sorted candidate paths that is able to establish the connection is allocated. For the confidential connections, the process of allocating spectrum resources is based on maximizing the number of XOR operations that can be performed for the selected path and group of frequency slots, utilizing one of the two aforementioned XOR-SSM metrics. Algorithm~\ref{alg1} below describes the NC-RSA approach for a given confidential demand. Also, an example in Fig.~\ref{temp_ex} is used to better explain the proposed algorithm when using either metric.

\begin{algorithm}
  \caption{NC-RSA for a given confidential demand} 
  \hspace*{\algorithmicindent} \textbf{Input:} $G(V,E),$ \textit{Set of candidate paths P}, \textit{Paths currently established $P_{e}$}, \textit{Confidential demand D ($s,d,B,1$)}, XOR threshold $T$, \textit{Spectrum slot metric} $\mathcal{S}$\\
  \hspace*{\algorithmicindent} \textbf{Output:} Confidential Connection Establishment 
  \label{alg1}
  \begin{algorithmic}[1]
  	\For{each candidate path $p \in P_{s,d}$}
  		\State Find a set of paths $P_{used} \in P_{e}$ that are used by the established connections, such that $\sum_{l=1}^{|V|} \delta_{p'l}^{p} > 0$, for each $p' \in P_{used}$
  	    \State Create temporary matrix $t$ = $L \times F$, where $L$ = number of links in path $p$ and $F$ = overall number of frequency slots.
  	    \State Initialize all entries of $t$ to $0$.
    	\For{each path $p_{u} \in P_{used}$}
    	    \State Using $\delta_{p_{u}}^{p}$ coefficient, increase by one the values for each link $l$ where $\delta_{p_{u}l}^{p}=1$ and spectrum slots that are covered by $p_{u}$ within $t$.
    	\EndFor
    	\State Initialize XOR-SSM$_{p} = 0$
    	\If {$\mathcal{S} == 0$} \Comment{MXOR-SSM is utilized} 
	\For{each group $(a-b)$ of available frequency slots (FS) in $p$ that can be allocated to $D$}
    	\State $c^{a-b}_{z}$= $\sum_{n=a}^{b} t_{z,n}$, for $z = 1$ to $L$ 
    	\State $\mathcal{C}^{a-b} = min(c^{a-b}_{1}, ..., c^{a-b}_{L})$
		\State XOR-SSM$_{p} = max($XOR-SSM$_{p}, \mathcal{C}^{a-b})$
	\EndFor	
	\Else   \Comment{AXOR-SSM is utilized}
	\For{each group $(a-b)$ of available FS in $p$ that can be allocated to $D$}
    	\State $c^{a-b}_{z}$= $\sum_{n=a}^{b} t_{z,n}$, for $z = 1$ to $L$ 
    	\If {$min(c^{a-b}_{1}, ..., c^{a-b}_{L}) \geq T$}
    	\State $\mathcal{C}^{a-b} =  (\sum_{z=1}^{L}c^{a-b}_{z})/L$
    	\State XOR-SSM$_{p} = max($XOR-SSM$_{p}, \mathcal{C}^{a-b})$
    	\EndIf
	\EndFor    	
    	\EndIf
    \EndFor
	\State Select candidate path $p$ with the highest value of the XOR-SSM metric.
	\If {XOR-SSM$_{p} \geq T$}
	\State Connection is \textbf{established} using path $p$ and group of frequency slots ($a-b$).
	\Else     
	\State Connection is \textbf{rejected}.
	\EndIf  
  \end{algorithmic}
\end{algorithm}

Specifically, in Algorithm~\ref{alg1}, for a given confidential connection, the XOR-SSM$_{p}$ metric is calculated for each candidate path $p$. To do this, a temporary matrix (denoted as $t$ in the algorithm) is used, with size $L \times F$, where $L$ denotes the number of links of the candidate path and $F$ denotes the overall number of frequency slots of each link in the network. Next, the set of the already established paths with a non-zero vector coefficient $\delta_{p_{u}}^{p}$  for path $p$ under investigation is found (denoted as $P_{used}$ in the algorithm, $p_u \in P_{used}$). This set of paths can be used to satisfy the ET constraint. Subsequently, for each path $p_u$, the values for the links and spectrum slots that are covered by $p_{u}$ within $t$ are increased by $1$. Next, for each available group of frequency slots $(a-b)$, the number of XOR operations performed on each link $z$ of path $p$ are calculated using the following equation: $c^{a-b}_{z}$ =  $\sum_{n=a}^{b}t_{z,n}$, where $a$ and $b$ are the starting and ending slots selected. In the case that MXOR-SSM is selected, the minimum $\mathcal{C}^{a-b}= min(c^{a-b}_{1}, ..., c^{a-b}_{L})$ value is selected (i.e., the weakest link in terms of the number of XOR operations performed in path $p$ is considered). On the other hand, in the case that AXOR-SSM is selected, then if the minimum number of XOR operations on each link is greater than or equal to the given acceptable threshold [$min(c^{a-b}_{1}, ..., c^{a-b}_{L}) \geq T$], the average number of XOR operations per link in the candidate path $p$ and group of frequency slots ($a-b$) is calculated via $\mathcal{C}^{a-b} =  (\sum_{z=1}^{L}c^{a-b}_{z})/L$. This procedure (for MXOR-SSM or AXOR-SSM) is repeated for all groups of frequency slots, and the group of frequency slots with the maximum value of $\mathcal{C}$ is selected. In addition, this procedure is repeated for all candidate paths that can establish connection $s$-$d$, and the maximum value of $\mathcal{C}$ amongst all candidate paths is selected  and is considered as the value of the XOR metric for path $p$ (using either the MXOR-SSM or the AXOR-SSM metric). If the XOR metric found has a value greater than or equal to threshold $T$, then the connection is established in that group of frequency slots. Otherwise, it is rejected.\\
\\
The value of the metric (e.g., $x$) selected signifies that either at least $x$ XOR operations are used to secure the confidential connection in each link (using MXOR-SSM) or an average number of $x$ XOR operations are used to provide security for each link of that connection, given that the threshold requirement $T$ is satisfied at each link of the path (using AXOR-SSM). Both processes also imply that the ET and FSM constraints are satisfied for that confidential connection. On the other hand, if $x < T$, there is at least one link in the path where the threshold requirement is not satisfied, and the connection is not considered secure.\\
\\
To clearly demonstrate this procedure, in Fig.~\ref{temp_ex}, assume a network with $F$ = $5$ and a candidate path $p_{1}$ ($1-2-6$) that requires $3$ frequency slots. Also, the set $P_{e}$ of connections currently established in the network consists of paths \{$p_{2}$, $p_{3}$, $p_{4}$, $p_{5}$, $p_{6}$, $p_{7}$, $p_{8}$\}. Table~\ref{pathinfo} provides information on each path (route along with the spectrum slots that each connection utilizes). Also, the $\delta$ coefficient is shown for each path in $P_{e}$ and path $p_1$.\\
\\
\\
\begin{table}[htbp]
\centering
\caption{Information for connections currently established in the network.}
\begin{tabular}{|c|c|c|c|c|}
\hline
\textbf{\begin{tabular}[c]{@{}c@{}}Connection\\ Id\end{tabular}} & \textbf{\begin{tabular}[c]{@{}c@{}}Path\\ Selected\end{tabular}} & \textbf{\begin{tabular}[c]{@{}c@{}}Spectrum\\ Slots allocated\end{tabular}} & $\delta_{p'1}^{p_{1}}$ & $\delta_{p'2}^{p_{1}}$ \\ \hline
$p_2$ & $2 - 4 - 6 - 3$ & $4 - 5$ & $0 $& $1$ \\ \hline
$p_3$ & $1 - 4 - 5 - 6$ & $3 - 4$ & $1$ & $1$ \\ \hline
$p_4$ & $2 - 3$ & $1 - 2$ & $0$ & $0$ \\ \hline
$p_5$ & $1 - 2$ & $1$ & $1$ & $0$ \\ \hline
$p_6$ & $6 - 2$ & $1 - 3$ & $0$ & $0$ \\ \hline
$p_7$ & $2 - 3 - 6$ & $4 - 5$ & $0$ & $1$ \\ \hline
$p_8$ & $1 - 4 -2$ & $1 - 2$ & $1$ & $0$ \\ \hline
\end{tabular}
\label{pathinfo}
\end{table}
\\
As discussed in the previous sections, a path must have at least two common nodes with the confidential one to provide security for that connection. Thus, $p_4$ is not able to provide any security for connection $p_1$. Also, since the direction of the path and the common nodes plays a key role in providing security for the confidential connection, path $p_6$ is also not able to provide any security for the given demand. As a result, $P_{used}$ consists of paths \{$p_{2}$, $p_{3}$, $p_{5}$, $p_{7}$, $p_{8}$\}. For example, connection $p_2$ contributes to cells $t_{(2,4)}$ and $t_{(2,5)}$, since it can cover spectrum slots $4$ and $5$ in the second link of $p_1$. On the other hand, connection $p_5$ could provide security to the confidential connection, but since it uses the same link as $p_1$, the spectrum slots allocated to $p_5$ (i.e., slot $1$) are excluded from the available group of frequency slots to satisfy the \textit{non-overlapping} constraint. This value is designated as `$-$'. The resulting $t$ matrix using all paths in $P_{used}$ is shown in Table~\ref{tab_slots}.

\begin{table}[htbp]
\parbox{.45\linewidth}{
\caption{Number of XOR operations per slot ($t$ matrix).}
\centering
\begin{tabular}{|c|c|c|c|c|c|c|}
\hline
\multicolumn{2}{|c|}{\multirow{2}{*}{\textbf{}}} & \multicolumn{5}{c|}{\textbf{Spectrum Slot id}} \\ \cline{3-7} 
\multicolumn{2}{|c|}{} & \textbf{1} & \textbf{2} & \textbf{3} & \textbf{4} & \textbf{5} \\ \hline
\multirow{2}{*}{\textbf{\begin{tabular}[c]{@{}c@{}}Link\\ id \end{tabular}}} & \textbf{1 (1-2)}&$-$&$1$&$1$&$1$&$0$\\ \cline{2-7} 
 & \textbf{2 (2-6)} &$-$&$0$&$1$&$3$&$2$\\ \hline
\end{tabular}
\label{tab_slots}
}
\hspace{0.6in}
\parbox{.4\linewidth}{
\caption{Number of XOR operations per link and group of available frequency slots.}
\centering
\begin{tabular}{|c|c|c|c|}
\hline
\multicolumn{2}{|c|}{} & $c_{z}^{2-4}$ & $c_{z}^{3-5}$ \\ \hline
\multirow{2}{*}{\textbf{\begin{tabular}[c]{@{}c@{}}Link\\ id (z)\end{tabular}}} & \textbf{1} & $3$ & $2$ \\ \cline{2-4} 
 & \textbf{2} & $4$ & $6$ \\ \hline
\end{tabular}
\label{XOR_link}
}
\end{table}

Next, the number of XOR operations must be calculated for each group of frequency slots that could be used to establish the confidential demand. To calculate the number of XOR operations on link $1$ and frequency slots $2 - 4$, $c^{2-4}_{1}$ = $t_{(1,2)}$ + $t_{(1,3)}$ + $t_{(1,4)}$ = $1$ + $1$ + $1$ = $3$. The same process is performed for all links that belong to the candidate path and the selected group of frequency slots $3-5$. It is noted that this process is not performed for spectrum slots $1-3$, since they are not available for the confidential demand. The result for each group of frequency slots and the links of the candidate path is shown in Table~\ref{XOR_link}.


Finally, to calculate the XOR-SSM metric, $\mathcal{C}$ must be calculated for each group of available frequency slots. In case the MXOR-SSM metric is used, $\mathcal{C}^{2-4}$ = $min(c^{2-4}_1,c^{2-4}_2)$ = $min(3,4)$= $3$, and $\mathcal{C}^{3-5}$ = $min(c^{3-5}_1,c^{3-5}_2)$ = $min(2,6)$ = $2$ respectively, as shown in Table~\ref{XOR_C1} for the group of frequency slots $2-4$ and $3-5$. Now, XOR-SSM$_{p_1}$ = $max(\mathcal{C}^{2-4}$, $\mathcal{C}^{3-5})$ = $max(3,2)= 3$, using frequency slots $2-4$. On the other hand, if the AXOR-SSM metric is used, $\mathcal{C}^{2-4}$ = $(c^{2-4}_1+c^{2-4}_2)/2$ = $(3+4)/2$ = $3.5$, and $\mathcal{C}^{3-5}$ = $(c^{3-5}_1+c^{3-5}_2)/2$ = $(2+6)/2$ = $4$ (Table~\ref{XOR_C2}). Hence, XOR-SSM$_p$ = $max(3.5,4)$ = $4$ using frequency slots $3-5$. It is noted that, for example, in the case $T$ = $3$, then this would not be an acceptable solution (i.e., $min(2,6)$ = $2 < T$) and therefore, frequency slots $2-4$ would be selected.

\begin{table}[htbp]
\parbox{.45\linewidth}{
\caption{Calculation of $\mathcal{C}$ for each group of frequency slots using MXOR-SSM.}
\centering
\begin{tabular}{|c|c|c|}
\hline
\multirow{3}{*}{} & \multicolumn{2}{c|}{\textbf{\begin{tabular}[c]{@{}c@{}}Group of \\ frequency \\ slots (x - y)\end{tabular}}} \\ \cline{2-3} 
 & \textbf{2 - 4} & \textbf{3 - 5} \\ \hline
$\mathcal{C}^{x-y}$ & $3$ & $2$ \\ \hline
\end{tabular}
\label{XOR_C1}
}
\hspace{0.4in}
\parbox{.45\linewidth}{
\caption{Calculation of $\mathcal{C}$ for each group of frequency slots using AXOR-SSM.}
\centering
\begin{tabular}{|c|c|c|c|}
\hline
\multirow{3}{*}{} & \multicolumn{2}{c|}{\textbf{\begin{tabular}[c]{@{}c@{}}Group of \\ frequency \\ slots (x - y)\end{tabular}}} \\ \cline{2-3} 
 & \textbf{2 - 4} & \textbf{3 - 5} \\ \hline
$\mathcal{C}^{x-y}$ & $3.5$ & $4$ \\ \hline
\end{tabular}
\label{XOR_C2}
}
\end{table}

\begin{figure}[htbp]
\centering
\includegraphics[scale = 0.4]{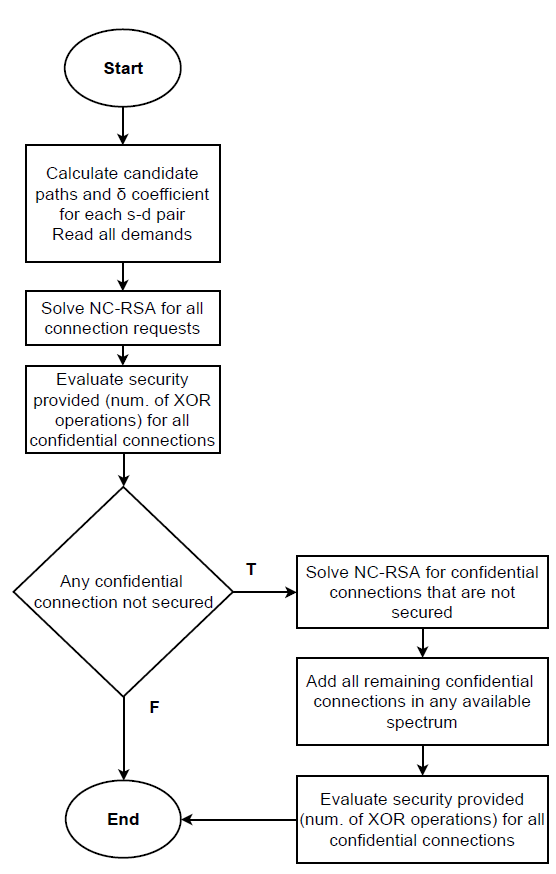}
\caption{Flowchart of proposed algorithm to minimize blocking (eliminate blocking due to the level of security requirements)}
\label{flowchart}
\end{figure}

Following either approach to solve the NC-RSA problem for the confidential demands will, in many cases, result in a high blocking probability, since  confidential connections that do not meet the required threshold $T$ are immediately blocked. To overcome this problem, the algorithms illustrated in the flowchart of Fig.~\ref{flowchart} is proposed.

As shown in the flowchart, the candidate paths are first pre-calculated and the $\delta$ coefficient is calculated for each pair of paths. Then, a typical RSA is implemented for the non-confidential demands with one of the routing strategies proposed, while the NC-RSA algorithm is used for the confidential demands with either the MXOR-SSM or AXOR-SSM metrics utilized. After all demands are allocated in the network, the confidential connections that are not secured (i.e., less XOR operations than the threshold $T$ are performed in at least one link of their selected path), and were previously rejected are re-entered in the network and the NC-RSA is again solved for these connections (utilizing now existing connections that were not available the first time the NC-RSA heuristic was implemented). Also, the new number of XOR operations is calculated for the already established demands. Finally, any confidential connection that was not previously established is the network, as it was not meeting the level of security requirements, is now established,  regardless of the level of security provided (these connections are later on in the performance section identified as unsecured confidential connections (e.g., Fig.~\ref{pSMUL} illustrates the percentage of secured confidential demands)). Nevertheless blocking still exists due to lack of network resources for both non-confidential as well as confidential connections. It is noted that additional (dummy) lightpaths can be added with the sole purpose to ensure security for the confidential connections that do not reach the required threshold $T$. However, this is not within the focus of this work and will be considered as future work.


\section{Performance Evaluation}
\label{results}
\begin{figure}[htbp]
\parbox{.35\linewidth}{
\includegraphics[scale = 0.14]{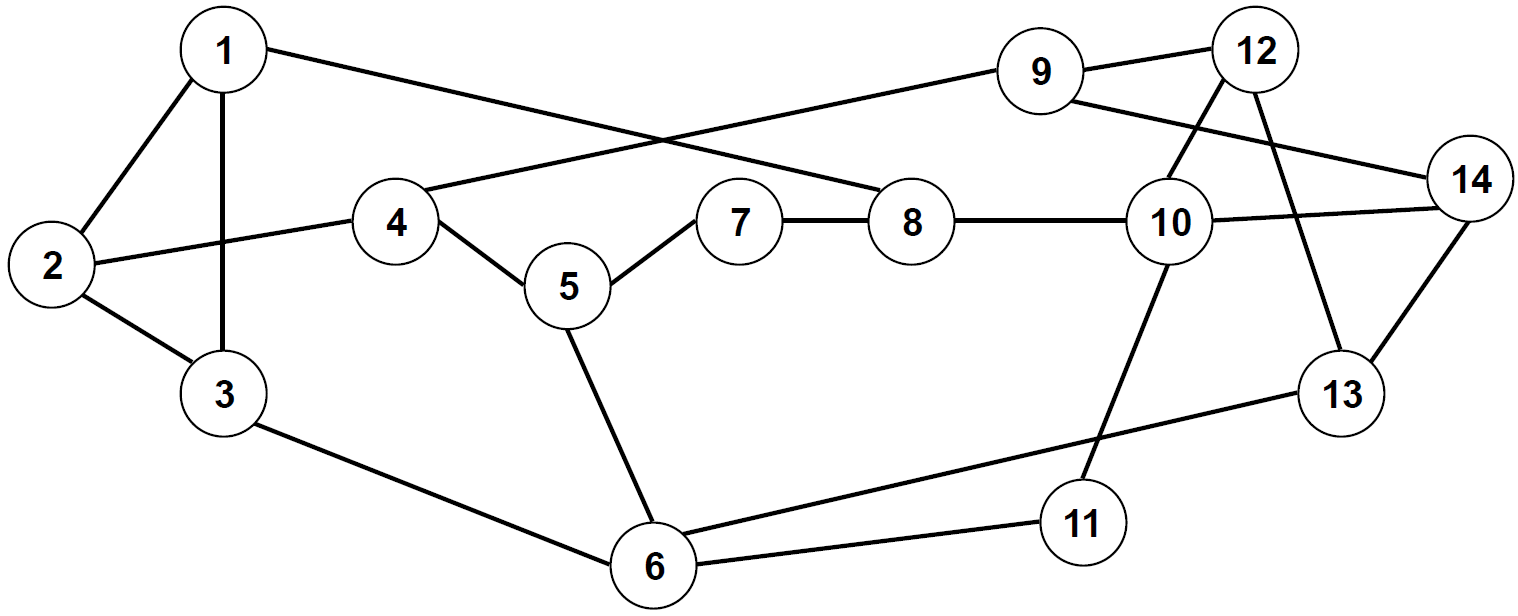}
\caption{Network topology used in the simulations, consisting of $14$ nodes and $42$ directed links.}
\label{Net}
}
\hspace{0.7in}
\parbox{.45\linewidth}{
\includegraphics[scale = 0.43]{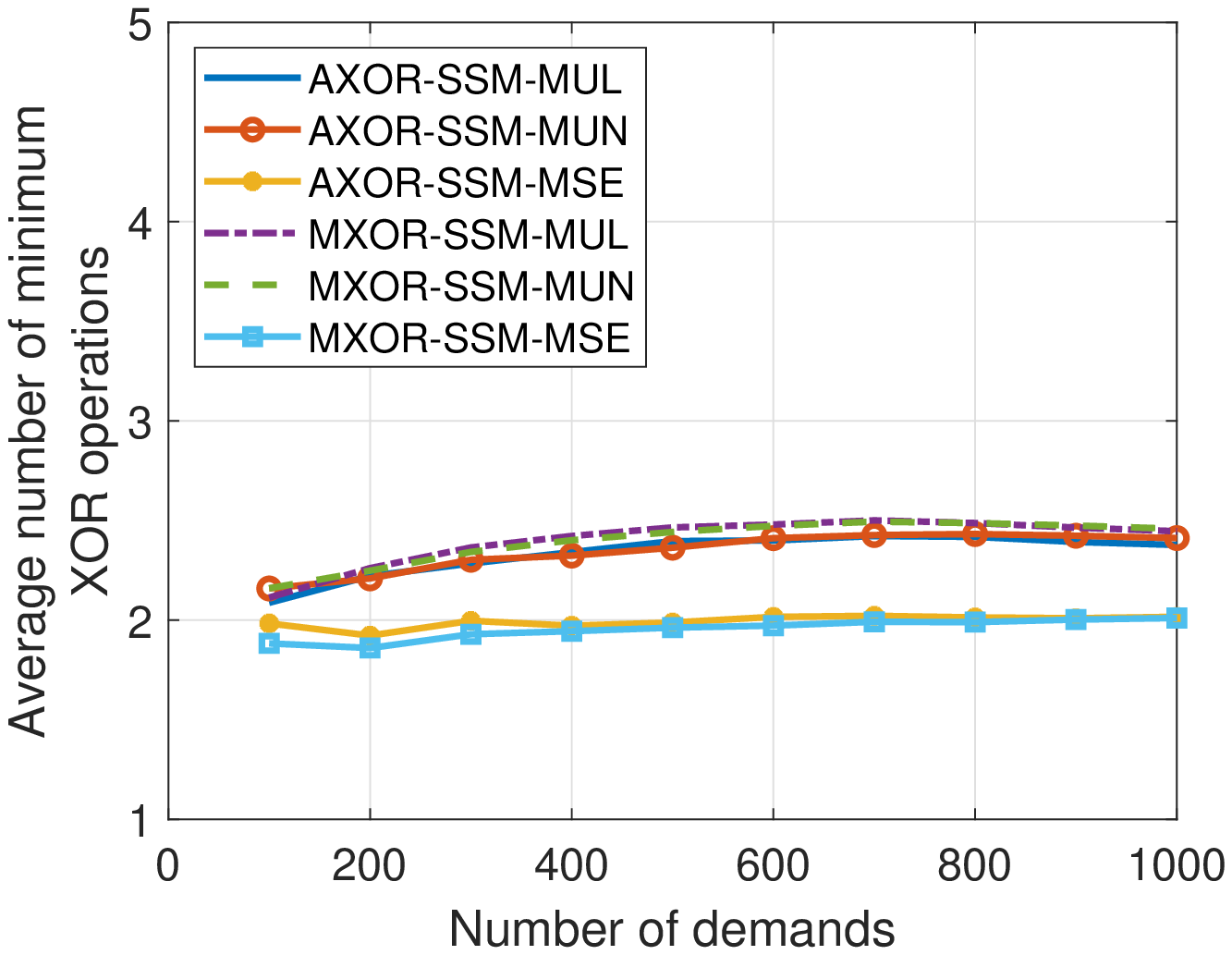}
\caption{Average number of minimum XOR operations over all confidential demands.}
\label{nutilnmxors5}
}
\end{figure}

To evaluate the proposed heuristic algorithm, an EON is implemented using bandwidth variable transponders that operate using the following modulation formats: BPSK, QPSK, $8$-QAM, and $16$-QAM. The transmission reach for each modulation format is given by $9300$, $4600$, $1700$, and $800$ km respectively. The network topology utilized in the simulations is shown in Fig.~\ref{Net}, consisting of $14$ nodes and $42$ directed links. For this network, a flexible grid is implemented, with a channel spacing of $12.5$ GHz which results in a total of $320$ spectrum slots with a baud rate of $10.7$ Gbauds for each link in the network. The requested demands are randomly generated using a uniform distribution for all $s$-$d$ pairs, where each demand size varies from $20$ to $100$ Gbps, while the number of candidate paths used for the first set of simulations is set to $5$ ($k$=$5$). Each presented result is the average of $10$ experiments performed with different generated sets of demands, where for all simulations $30\%$ of the overall number of demands are designated as confidential. Finally, a PC with an i7-3930K CPU and $24$ GB RAM is used for all simulations.

When using network coding, the level of security provided for a confidential connection can by quantified by the minimum and average number of XOR operations performed on each link of its established path. In this work, the threshold for the number of XOR operations that a connection must have on each link of its path in order to be considered as secure is set to $1$ ($T=1$). This means that the number of XOR operations ($N$) that must be performed at each link of the path used by a given confidential connection must be $N \geq 1$, in order to consider the specific connection as secure. 

By increasing the minimum number of XOR operations that occur at the links of the established path for the confidential connection, then even if an eavesdropper taps the weakest link of this confidential connection (i.e., the link with the smallest number of XOR operations performed), all connections used in these XOR operations will have to be compromised before the eavesdropper can make sense of the confidential information. The average number of minimum XOR operations over all confidential demands is presented in Fig.~\ref{nutilnmxors5}, when the three routing strategies are utilized for both the AXOR-SSM and MXOR-SSM metrics.

Using the MXOR-SSM-MUL or MXOR-SSM-MUN routing strategies provides the highest minimum number of XOR operations performed for each demand, on average, compared to the rest of the cases examined. This is to be expected, since the MXOR-SSM metric has as an aim to select the candidate path and group of frequency slots that maximize the minimum number of XOR operations for the confidential demand. Further, it is shown that by using the AXOR-SSM metric and either the MUL or MUN routing algorithms provides results close to the MXOR-SSM algorithm. This is due to the fact that in this case a relatively small number of candidate paths is utilized. Thus, in order to satisfy the required XOR threshold for each link, while also maximizing the average number of XOR operations, a solution where the minimum number of XOR operations is close to the MXOR-SSM case is selected. Indeed, as shown in Table \ref{tab:dif_k} below, the difference between AXOR-SSM and MXOR-SSM, in terms of the minimum number of XOR operations per demand, increases with the number of candidate paths. Finally, the MSE routing strategy provides the least number of XOR operations per confidential connection which is to be expected, since it has as an aim to add non-confidential connections in a spectrum efficient manner, rather than to provide more XOR operations for the confidential ones. Nevertheless, in all cases, the results indicate that on the average the minimum number of XOR operations that take place per link per demand is between $2$-$3$, higher than the predefined XOR threshold ($T=1$).

\begin{figure}[htbp]
\parbox{.45\linewidth}{
\includegraphics[scale = 0.43]{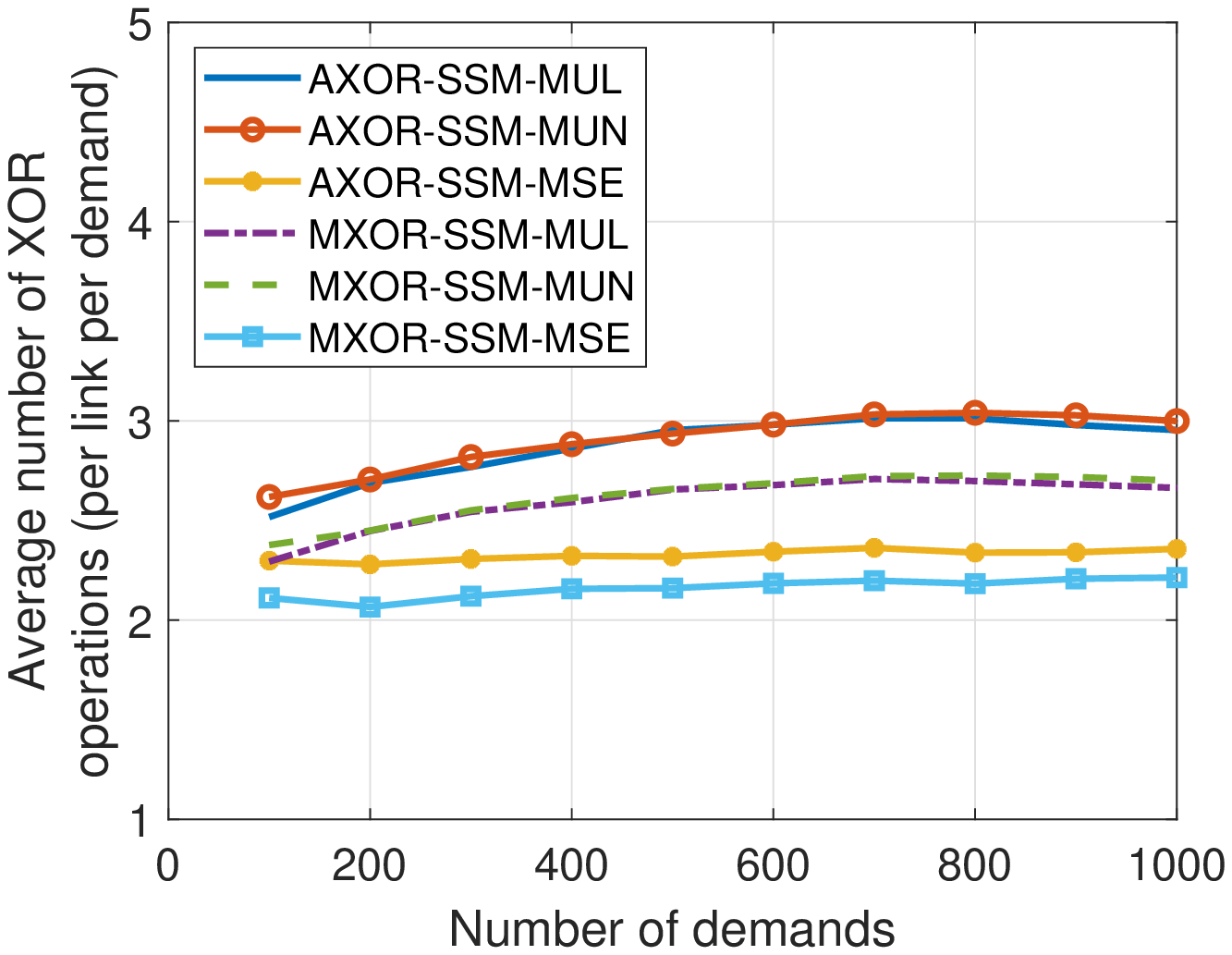}
\caption{Average number of XOR operations per link per confidential connection.}
\label{nutilnxors5}
}
\hspace{0.2in}
\parbox{.45\linewidth}{
\centering
\includegraphics[scale = 0.43]{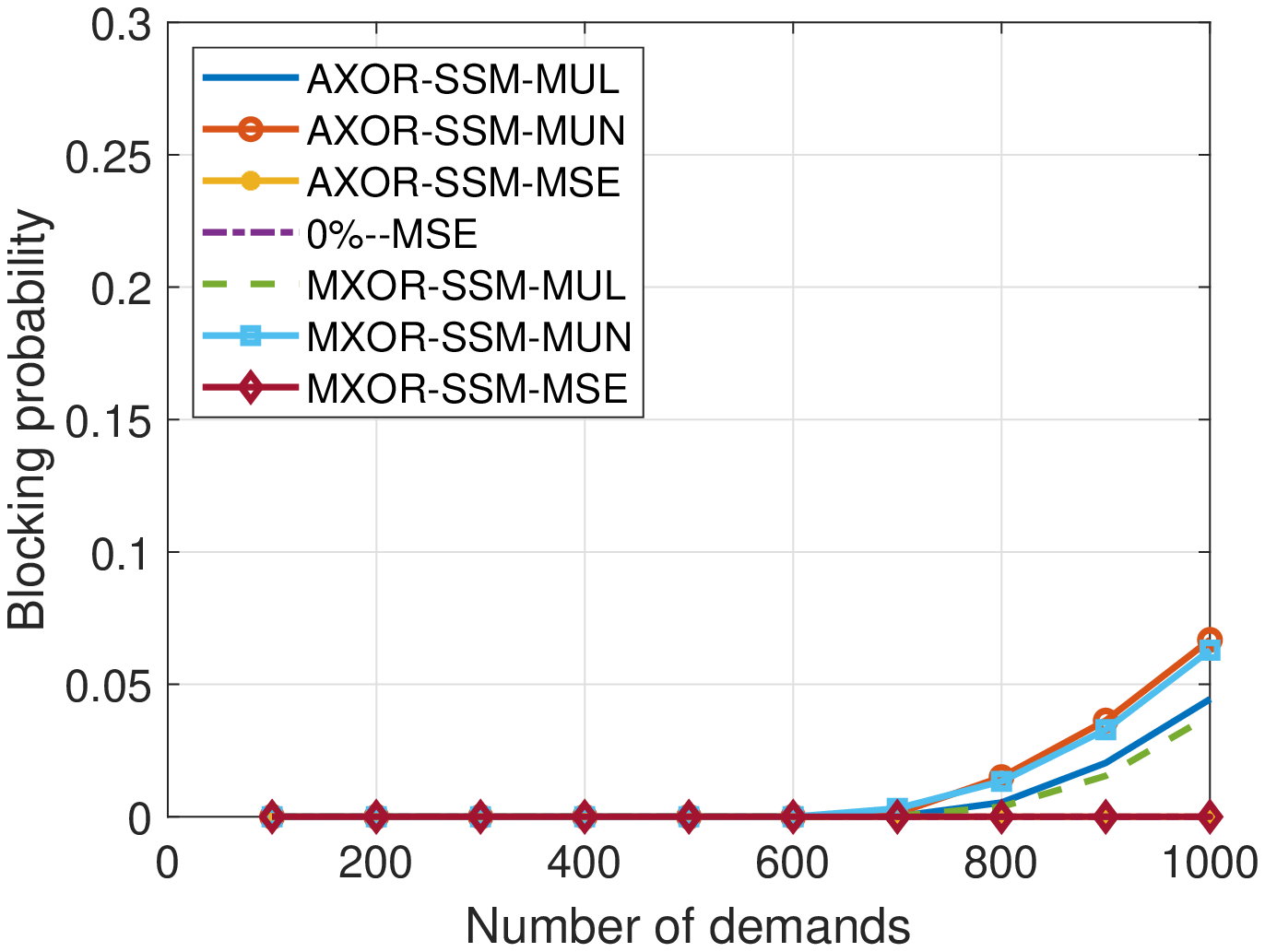}
\caption{Blocking probability for different routing strategies and XOR spectrum slot metrics.}
\label{net_blocknew}
}
\end{figure}

As discussed in the previous sections, the number of XOR operations performed on each link is a key point for the level of security provided to a confidential demand. Accounting for the average number of XOR operations per link per demand is another useful metric that could describe the average amount of connections that have to be compromised to make the confidential demand vulnerable. Thus, Figure~\ref{nutilnxors5} presents the average number of XOR operations per link per confidential connection in the network, when using different slot allocation metrics and routing techniques for the candidate paths. 


The AXOR-SSM metric provides the best results in terms of the average number of XOR operations performed, which is to be expected, since it has as an aim to maximize the total number of XOR operations for a connection, as long as the XOR threshold is satisfied for all links. On the other hand, using the MXOR-SSM metric, which aims to maximize the minimum number of XOR operations over all links of the demand, does not always provide the best results in terms of the total number of XOR operations performed for the confidential connections. Further, the MUL routing strategy results in a larger number of XOR operations compared to both the MSE and MUN routing strategies when either spectrum allocation metric is used. This is due to the fact that since the direction of the paths in the XOR process is taken into consideration, a confidential path that can use other existing connections traversing a highly utilized link (i.e., traverses the nodes in the same order through a link-disjoint path) will potentially increase the XOR operations performed in its path. Thus, utilizing the MUL sorting technique maximizes the utilization of specific links in the network, which subsequently results in providing a higher level of security for the confidential demands (contrary to results in~\cite{Savva19}, where the direction of the paths was not considered).

Next, the blocking probability of the network is presented (Fig.~\ref{net_blocknew}) when different metrics and routing techniques are utilized. In this figure, the case where all connections are designated as non-confidential, allocated using the MSE routing strategy, is also presented as a benchmark (designated as ``0\% MSE''). It is clearly demonstrated that the MUL routing strategy results in a much higher blocking probability compared to the MSE case, since the paths chosen, utilizing the former strategy, aim to increase the level of security of the confidential connection (by being  utilized at the most used links or nodes), rather than to maximize the efficient use of spectrum resources, which is precisely the aim of the latter strategy. In fact, the MSE approach, satisfies the appropriate threshold of XOR operations per link in the network, while also providing a very low blocking probability, similar to the performance of the benchmark case.


Clearly, the number of candidate paths for each $s$-$d$ pair that a confidential connection can use will also have a great impact on the network performance in terms of the level of security, blocking probability, and spectrum utilization, as shown in Table~\ref{tab:dif_k}. Note that only results for the MXOR-SSM-MUL, AXOR-SSM-MUL and AXOR-SSM-MSE policies are shown,  as these policies provide the best results (as previously shown in Figs.~\ref{nutilnmxors5}-\ref{net_blocknew}). Again, the results for the 0\%-MSE case are provided as benchmark.

\begin{table}[htbp]
\centering
\caption{Impact of $k$ on level of security, blocking probability, and spectrum utilization.}
\begin{tabular}{|c|c|c|c|c|c|}
\hline
\multirow{2}{*}{\textbf{Result}} & \multirow{2}{*}{\textbf{Technique}} & \multicolumn{3}{c|}{\textbf{k}} \\ \cline{3-5} 
 &  & \textbf{5} & \textbf{10} & \textbf{20} \\ \hline
\multirow{4}{*}{\textbf{\begin{tabular}[c]{@{}c@{}}Average number of\\ minimum XOR\\  operations\end{tabular}}} & \textbf{MXOR-SSM-MUL} & 2.45 & 2.83 & 3.09 \\ \cline{2-5} 
 & \textbf{AXOR-SSM-MUL} & 2.38 & 2.75 & 2.99 \\ \cline{2-5} 
 & \textbf{AXOR-SSM-MSE} & 2.02 & 2.19 & 2.50 \\ \cline{2-5} 
 & \textbf{0\% - MSE} & - & - & - \\ \hline
\multirow{4}{*}{\textbf{\begin{tabular}[c]{@{}c@{}}Average number\\ of XOR operations\\ per link per demand\end{tabular}}} & \textbf{MXOR-SSM-MUL} & 2.66 & 3.17 & 3.60 \\ \cline{2-5} 
 & \textbf{AXOR-SSM-MUL} & 2.95 & 3.76 & 4.28 \\ \cline{2-5} 
 & \textbf{AXOR-SSM-MSE} & 2.36 & 2.87 & 3.57 \\ \cline{2-5} 
 & \textbf{0\% - MSE} & - & - & - \\ \hline
\multirow{4}{*}{\textbf{\begin{tabular}[c]{@{}c@{}}Blocking\\ probability\end{tabular}}} & \textbf{MXOR-SSM-MUL} & 0.04 & 0.11 & 0.17  \\ \cline{2-5} 
 & \textbf{AXOR-SSM-MUL} & 0.04 & 0.13 & 0.20 \\ \cline{2-5} 
 & \textbf{AXOR-SSM-MSE} & 0 & 0 & 0 \\ \cline{2-5}  
 & \textbf{0\% - MSE} & 0 & 0 & 0 \\ \hline
\multirow{4}{*}{\textbf{\begin{tabular}[c]{@{}c@{}}Number of\\ spectrum slots\\ utilized\end{tabular}}} & \textbf{MXOR-SSM-MUL} & 7561 & 8429 & 9219 \\ \cline{2-5} 
 & \textbf{AXOR-SSM-MUL} & 7734 & 8723 & 9456 \\ \cline{2-5}
 & \textbf{AXOR-SSM-MSE} & 5243 & 5754 & 6538 \\ \cline{2-5}  
 & \textbf{0\% - MSE} & 4232 & 4257 & 4322 \\ \hline
\end{tabular}
\label{tab:dif_k}
\end{table}

From Table~\ref{tab:dif_k}, it is evident that the larger the number of candidate paths for each $s$-$d$ pair, the larger the number of XOR operations that can be performed for each confidential demand. Thus, both the average and minimum number of XOR operations increase as $k$ increases, for all NC-RSA strategies. It should be noted though, that  as $k$ increases, the number of additional spectrum slots utilized, as well as the blocking probability will also increase, since now both confidential and non-confidential demands will use paths that are not efficient in terms of spectrum utilization, as they instead select a solution which maximizes the number of XOR operations performed by the confidential demands. Thus, the choice of $k$ should be one that provides a balanced solution between the efficiency of the network and the level of security provided for the confidential demands.
 
From the results presented, for any choice of $k$ (and as discussed previously), the AXOR-SSM-MUL approach provides the highest number of XOR operations on average for each confidential demand, while also satisfying the minimum threshold for each link for all confidential connections. On the other hand, the AXOR-SSM-MSE approach provides a much lower blocking probability, while achieving a slightly smaller number of XOR operations. Thus, for the rest of the simulations, the AXOR-SSM-MUL approach, which provides the best results in terms of the level of security provided for the confidential demands and the AXOR-SSM-MSE approach, which provides the most efficient usage of the spectrum while providing an acceptable level of security for the confidential connections will be utilized. In conjunction, a set of $10$ candidate paths for each $s$-$d$ pair will be used, in order to provide a balanced solution in terms of level of security and spectrum efficiency. Also, $1500$ connections are now established in the network to better illustrate how each approach performs as the number of demands increases considerably.

\begin{figure}[htbp]
\centering 
\includegraphics[scale = 0.45]{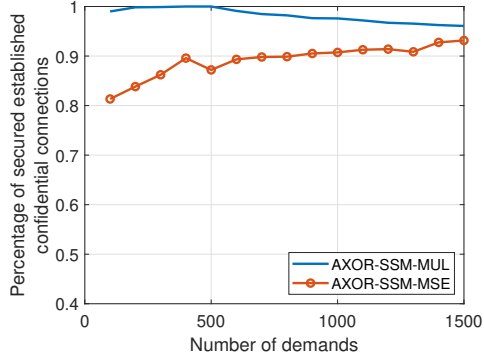}
\caption{Percentage of secured confidential connections established in the network ($k=10$).}
\label{pSMUL}
\end{figure}

The percentage of the established confidential connections that are properly secured (i.e., satisfying the \textbf{ET} and \textbf{FSM} constraints) is subsequently presented in Fig.~\ref{pSMUL}. As shown in this figure, different routing techniques can have a significant effect on the number of secure confidential connections. Clearly, from the results presented, the MUL technique performs better compared to the MSE routing strategy, since connections are forced to traverse the same links, and therefore, more connections can be found that can be XOR-ed with the confidential connection. However, it is noted that when using the MSE routing strategy, the percentage of secure connections increases with the number of demands that are established in the network. On the other hand, the percentage of confidential connections that are securely established in the network utilizing the MUL strategy decreases as the number of demands increases, since the resources of the candidate paths that could provide security for the confidential demands are saturated, leading to the usage of less secure candidate paths. In order to secure all confidential demands, it is again noted that additional dummy lightpaths can be utilized in the XOR process in order to provide security for any unsecured confidential connections established in the network.

\begin{figure}[htbp]
\parbox{.50\linewidth}{
\centerline{\includegraphics[scale = 0.43]{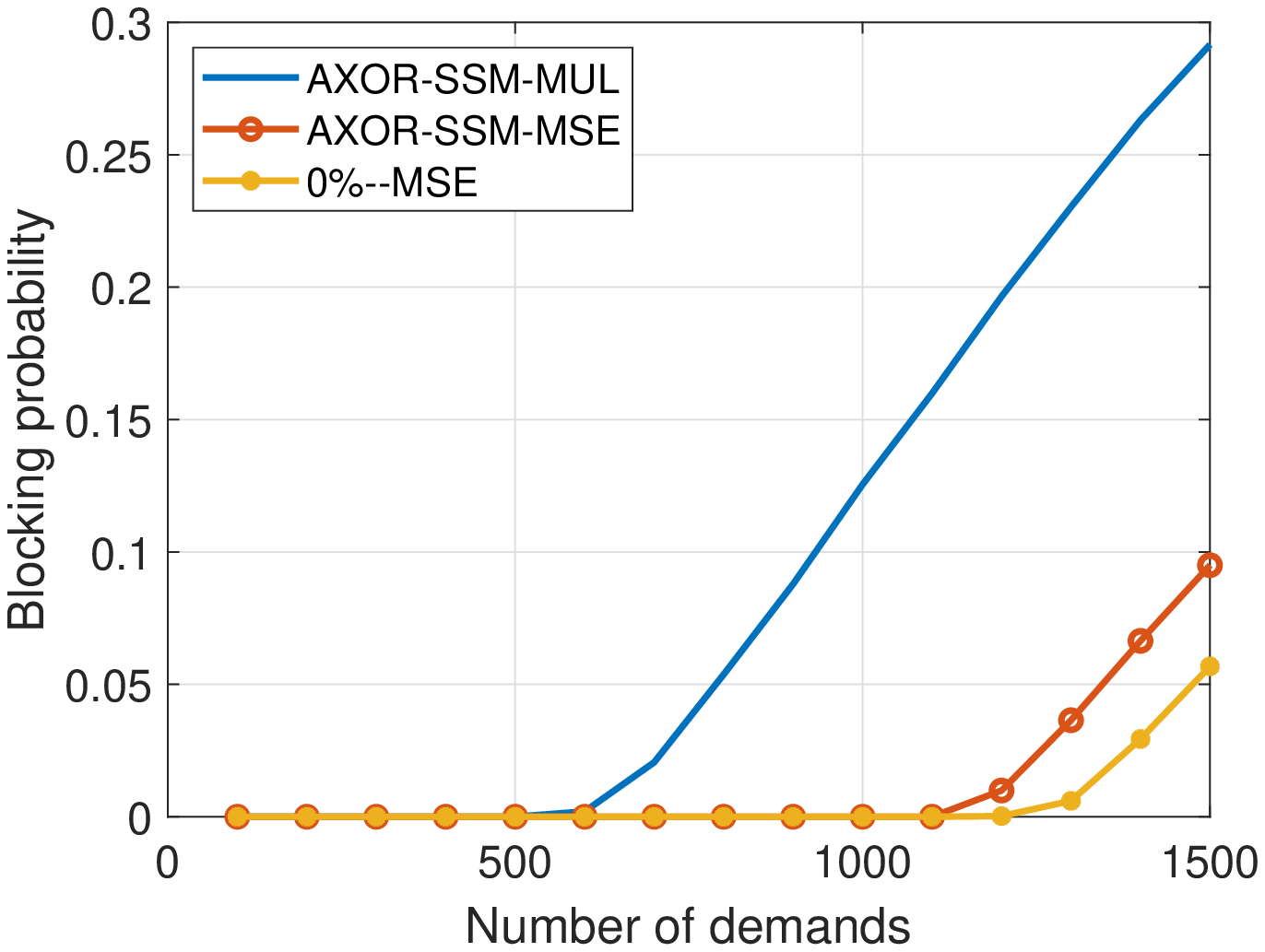}}
\caption{Blocking probability using \newline different routing techniques ($k=10$).}
\label{block_fig}
}
\hfill
\parbox{.50\linewidth}{
\centerline{\includegraphics[scale = 0.43]{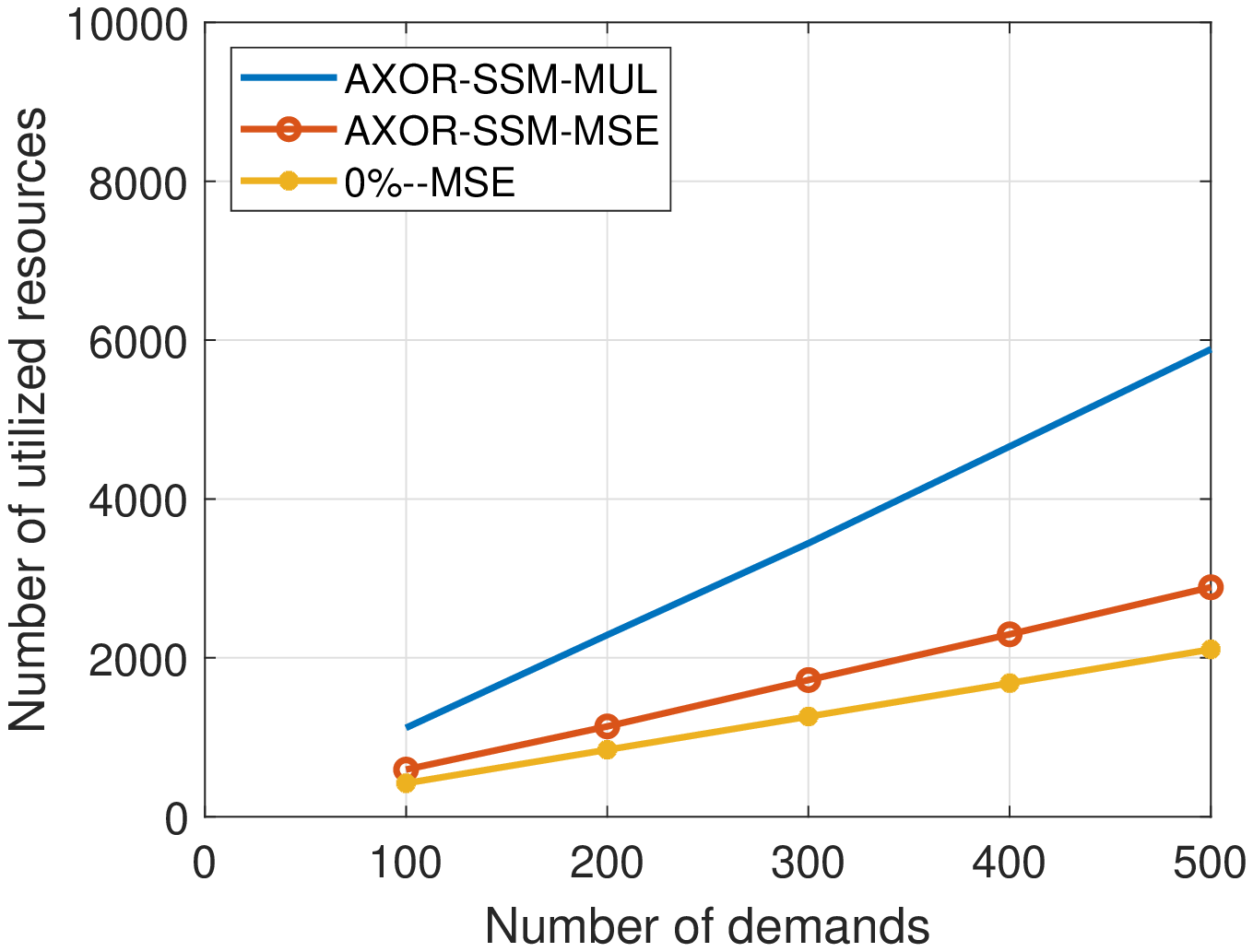}}
\caption{Number of spectrum slots utilized to establish all connections in the network.}
\label{utili_res}
}
\end{figure}

Next, the blocking probability is presented in Fig.~\ref{block_fig} when more connections ($1500$) are established in the network and more candidate paths ($k=10$) are considered for each connection. From the results obtained, again the MUL sorting strategy is shown to provide a much higher blocking probability compared to the MSE strategy. In fact, the MSE strategy provides a much lower blocking probability, close to the benchmark case, for a large percentage of connections (more than $90\%$ for more than $1000$ demands), while also satisfying the appropriate threshold of XOR operations per link in the network (Fig.~\ref{pSMUL}). 


Selecting different routing and spectrum allocation strategies for each confidential connection will, on the one hand,  secure the confidential connections against an eavesdropping attack, but, on the other hand, could force the algorithm to deviate from an efficient spectrum utilization solution. Figure~\ref{utili_res} presents the number of utilized spectrum slots when using the MUL and MSE routing strategies versus the benchmark case (all connections are designated as non-confidential and the candidate paths are routed based on the MSE technique). In this case, there is no blocking in the network (enough network resources are available), in order to obtain the exact number of spectrum resources utilized for each set of demands. As shown in the figure, the MUL routing strategy (which provides the largest number of XOR operations, on average, for the confidential connections) requires much more spectrum resources to establish all connections, compared to MSE. On the other hand, using the MSE approach to provide security for confidential connections results in an acceptable (i.e., above designated threshold) number of XOR operations on average for the confidential demands, while at the same time it requires much fewer resources compared to the MUL case. In fact, using MSE, the number of additional resources allocated to the network is slightly increased compared to the case where all connections are not confidential (``0\%-MSE'' - benchmark case). This is the case, since all non-confidential connections are assigned using the best path available (in terms of hops and modulation format).

\vspace{-0.3in}

\section{Conclusions}
\label{conclusion}
In this work, a novel heuristic approach is proposed, that utilizes the concept of network coding in EONs to provide security for confidential connections against eavesdropping attacks. Using the proposed approach, the physical layer security is increased, since connections established at link-disjoint paths combine their datastreams to encrypt the confidential data, thus an eavesdropper must now compromise several connections traversing different parts of the network in order to make sense of any accessed confidential data. 

Performance results indicate that using the average XOR spectrum slot metric (AXOR-SSM) to evaluate each candidate path and group of spectrum slots required for the confidential demands results in a higher number of XOR operations performed on each link of the confidential path, and thus it provides an increased level of security against an adversary that attempts to compromise confidential data. Also, the combination of the AXOR-SSM metric with the most used links (MUL) routing strategy, provides the highest number of XOR operations for each confidential demand at the expense of a significant increase in the spectrum resources utilized, while the maximum spectrum efficiency (MSE) routing strategy provides a similar number of XOR operations at the expense of a small percentage of additional spectrum slots. Therefore, using the AXOR-SSM-MSE approach, the required level of security of the confidential connections is achieved, while spectrum utilization and blocking probability are similar to the ones for the benchmark case, where all connections are considered as non-confidential.

Ongoing work includes the development of integer linear programming (ILP) formulations to provide an optimal solution to the NC-RSA problem, as well as the consideration of additional physical layer constraints that are imposed by the utilization of network coding via XOR operations.



\end{document}